\documentclass[10pt, conference, letterpaper]{IEEEtran}
\IEEEoverridecommandlockouts

\usepackage{amsmath}
\usepackage{amssymb}
\usepackage{mathrsfs}
\usepackage{amsfonts}
\usepackage[linesnumbered, ruled]{algorithm2e}

\usepackage{cite}
\usepackage{multirow}
\usepackage{tabularx}
\usepackage{amsthm}
\usepackage{algorithmic}
\usepackage{graphicx}
\usepackage{multicol}
\usepackage{subfigure}
\usepackage{bm}
\usepackage{booktabs}

\begin{document}
{\setlength\abovedisplayskip{1pt}
\setlength\belowdisplayskip{1pt}}


\title{Interference-aware User Grouping Strategy in NOMA Systems with QoS Constraints}

\author{\IEEEauthorblockN{Fengqian Guo, Hancheng Lu, Daren Zhu, Hao Wu}
   \IEEEauthorblockA{
       The signal Network Lab of EEIS Department, USTC, Hefei, China, 230027 \\
		Email: fqguo@mail.ustc.edu.cn, hclu@ustc.edu.cn, darenzhu@mail.ustc.edu.cn, hwu2014@mail.ustc.edu.cn
	}
}

\maketitle

\begin{abstract}
To meet the performance and complexity requirements from practical deployment of non-orthogonal multiple access (NOMA) systems, several users are grouped together for NOMA transmission while orthogonal resources are allocated among groups. User grouping strategies have significant impact on the power consumption and system performance. However, existing related studies divide users into groups based on channel conditions, where diverse quality of service (QoS) and interference have not been considered. In this paper, we focus on the interference-aware user grouping strategy in NOMA systems, aiming at minimizing power consumption with QoS constraints. We define a power consumption and externality (PCE) function for each user to represent the power consumption involved by this user to satisfy its QoS requirement as well as interference that this user brings to others in the same group. Then, we extend the definition of PCE to multi-user scenarios and convert the user grouping problem into the problem of searching for specific negative loops in the graph. Bellman-Ford algorithm is extended to find these negative loops. Furthermore, a greedy suboptimal algorithm is proposed to approach the solution within polynomial time. Simulation results show that the proposed algorithms can considerably reduce the total power consumption compared with existing strategies.
\end{abstract}


\newtheorem{def1}{\bf Definition}
\newtheorem{thm1}{\bf Theorem}
\newtheorem{lem1}{\bf Lemma}
\newtheorem{cor1}{\bf Corollary}

\section{Introduction}

With the rapid development of mobile communications, 5G faces arising challenges due to the sparser spectrum resources. To improve the spectrum efficiency,  non-orthogonal multiple access (NOMA) has been considered as a promising technology \cite{Dai1, Chang2018, 8003496, Zeng2018}. Different from conventional orthogonal multiple access (OMA), NOMA allows multiple users to share the same time/frequency resource blocks by superposition coding (SC)\cite{Islam}, and uses different power levels to distinguish the signals of different users. Then the receivers apply successive interference cancellation (SIC) for multi-user detection and decoding \cite{Sai1}.

Although NOMA can improve the spectrum efficiency by multiplexing signals of different users in the power domain, interference among users is severe. This would degrade the decoding performance and increase SIC decoding complexity, when NOMA is implemented among all users simultaneously\cite{Wei3}. Therefore, in general, users are divided into different groups, then NOMA is implemented within each group and traditional OMA is carried out among different groups. The process of dividing users into different groups is called user grouping. Different user grouping strategies result in different power consumptions. Hence it's important to find an appropriate user grouping strategy to reduce power consumption as well as enlarge the performance gain of NOMA \cite{dzg1, Cui2018}. On the other hand, a user will interfere with the other users in the same group, i.e., the externalities. Such kind of interference has signifciant impact on the performance and power consumption of NOMA systems. Consequently, when grouping a user, both the power consumption of this user and its externalities should be considered, which makes the user grouping problem much more intractable.

As User grouping is critical for practical deployment of NOMA systems, many research attempts have been made on this issue \cite{dzg1, Lv2018, Ali1, Kiani2018, Zhou2018, Han2018}. For instance, Ding \emph{et al}. in \cite{dzg1} proposed that selecting two users whose channel conditions are more distinctive into the same group, which can increase the sum-rate of NOMA systems. The authors in \cite{Ali1} expanded this strategy to a scenario where more users are assigned into the same group. There also exist some user grouping strategies based on the maching theory or matching game \cite{Liang2017}, where the set of users and channels were considered as two sets of players \cite{Marcano2018, zhang1, Xu2018, Di1}. The authors in \cite{Xu2018} proposed a suboptimal algorithm based on matching theory to maximize the system capacity in half-duplex cognitive OFDM-NOMA systems. In \cite{Di1}, users were assigned one by one based on matching game to maximize the weighted total sum-rate with consideration of user fairness.

However, there still exist challenges that obstruct existing user grouping strategies from achieving the optimal performance, especially in power consumption. In practice, different users have different requirements on quality of service (QoS), usually in terms of data rate. Such diverse QoS requirements have not been considered in existing user grouping strategies, while they have significant impact on power allocation and user grouping. On the one hand, sufficient power should be allocated to users to provide them with target data rates. On the other hand, users with higher target data rates will produce more severe interference to others in the same group when NOMA is applied. From the view point of power efficiency, assigning users with high target data rates to the same group should be avoided as much as possible. Based on the above analysis, we can see that power allocation and user grouping are coupled with each other when QoS requirements of users are considered. To achieve the best performance, they should be jointly optimized. Unfortunately, this joint optimization problem is non-trivial due to intra-group interference among users. Such kind of interference makes users interrelated in power allocation and group selection. On the contrary, most existing user grouping strategies are based on channel conditions of individual users, the impact of interference among users is not well studied. Furthermore, existing user grouping strategies are performed under the assumption that the number of users in each group is equal, which might not be optimal. When the number of users in each group can be arbitrary, the user grouping problem will become much more challenging.

In this paper, to address aforementioned challenges, we study the interference-aware user grouping strategy in NOMA systems, aiming at minimizing power consumption with QoS constraints. The main contributions are described as follows.

\begin{itemize}
\item We define a power consumption and externality (PCE) function for each user with QoS requirement, which  represents the power consumption involved by this user as well interference that this user brings to others in the same group, i.e., externalities in NOMA. We prove that if a user changes its grouping strategy, the difference of its PCE function will match the difference of the total power consumption exactly.

\item To extend the property of the PCE function to multi-user scenarios, \emph{shift league} and \emph{exchange league} are defined, which are used to judge whether the total power consumption can be reduced by changing the grouping strategies of the users in different groups. Based on these definitions, we construct a directed graph and convert the user grouping problem into the problem of searching for specific negative loops in the graph.

\item We extend Bellman-Ford algorithm to find the negative loops with all users in different groups. Furthermore, to reduce the computational complexity, we propose a fast suboptimal algorithm based the greedy strategy to approach the solution within polynomial time.

\end{itemize}

Extensive simulations demonstrate that the proposed algorithms outperform existing user grouping strategies in terms of power consumption under different scenarios and converge to a stable solution in finite iterations.

The rest of this paper is organized as follows. Section II outlines the model of the NOMA system with user grouping. The user grouping problem with QoS constraints is analyzed and formulated in Section III. In Section IV, the PCE function is defined for each user to measure the power consumption and interference introduced by this user. In Section V, we convert the user grouping problem into the problem of searching for specific negative loops in the graph and propose two algorithms to solve it. System performance is evaluated in Section VI. Finally, conclusion is drawn in Section VII.

\emph{Notations:} Vectors and sets are denoted by bold and calligraphic letters, respectively. $\cup$ and $\setminus$ represent set union and set difference operators, respectively.
$\lceil \cdot\rceil $ denotes the ceiling function. Given a set $\mathcal{A}$, $\|\mathcal{A}\|$ denotes the number of elements in $\mathcal{A}$.
Table I lists the key notations used in this paper.

\begin{table}
	\centering
	\caption{Key Notations}
	\begin{tabular} {m{40pt}	m{183pt}	m{0cm}}
		\toprule
		Notation		& Meaning							 	\\
		\midrule
		$ \mathcal{N} $ 		& Index set of users 										\\		
		$ \mathcal{G} $ 		& Index set of groups 									\\		
		$ \mathcal{U}^g $ 		& Index set of users in group $ g $ 							\\
$ p_n $ 			& Transmit power allocated to user $ n $ 	\\
		$ p^g_k $ 			& Transmit power allocated to the $ k $-th user in group $ g $ 	\\
		$ P_t $ 				& Total transmit power allocated to all users					\\
		$ P^g $ 				& Total transmit power allocated to the users in group $ g $		\\
		$ \pi_n $ 				& Group that user $n$ assigned to 						\\
		$ \bm{\pi} $ 			& Vector of grouping strategy of all users 			\\
		$ \bm{\pi}_{-n} $ 		& Vector of grouping strategy of users except user $ n $ 	\\
		$ \varepsilon^g_{i,j} $ 	& Power consumption that the $ i $-th user in group $ g $ brings to the $ j $-th user in the same group \\
		$ E_{n} $ 	& Externalities function of user $n $ 	\\
        	$ \mathcal{C}_{n} $ 		& PCE function of user $ n $ \\
       		$ \!\!\!(n\!\in\!\mathcal{U}^g\!,\!\bm{\pi}_{\!-\!n}\!) $ 	& User $ n $ is assigned into group $ g $ and the grouping strategy of the other users is $ \bm{\pi}_{-n} $	\\
        	$ (n\!\!\rightarrow\!\! g) $ 	& Operation that indicate user $ n $ is moved into group $ g $					\\
        	$ (n\!\!\twoheadrightarrow\!\!\tilde{n}) $ & Operation that indicate user $ n $ is moved into the group including user $ \tilde{n} $, and user $ \tilde{n} $ is moved out of group $ \pi_{\tilde{n}} $\\
		\bottomrule
	\end{tabular}
	\label{table1}
\end{table}


\section{System Model}

Consider a downlink NOMA systems with one base station (BS) and $N$ single-antenna users. The $N$ users are divided into $G$ groups. The number of groups is up to the number of available channels. Assuming that channel state information (CSI) is available at users and BS.
Let $\mathcal{N}=\{1, \cdots , N\} $ denote set of the indexes of users, and $\mathcal{G}= \{1,\cdots, G\}$ denote set of the indexes of groups. NOMA is implemented within each group and traditional OMA is carried out among different groups.
\begin{figure}[htbp]
	\centering
	\includegraphics[width=3.49in]{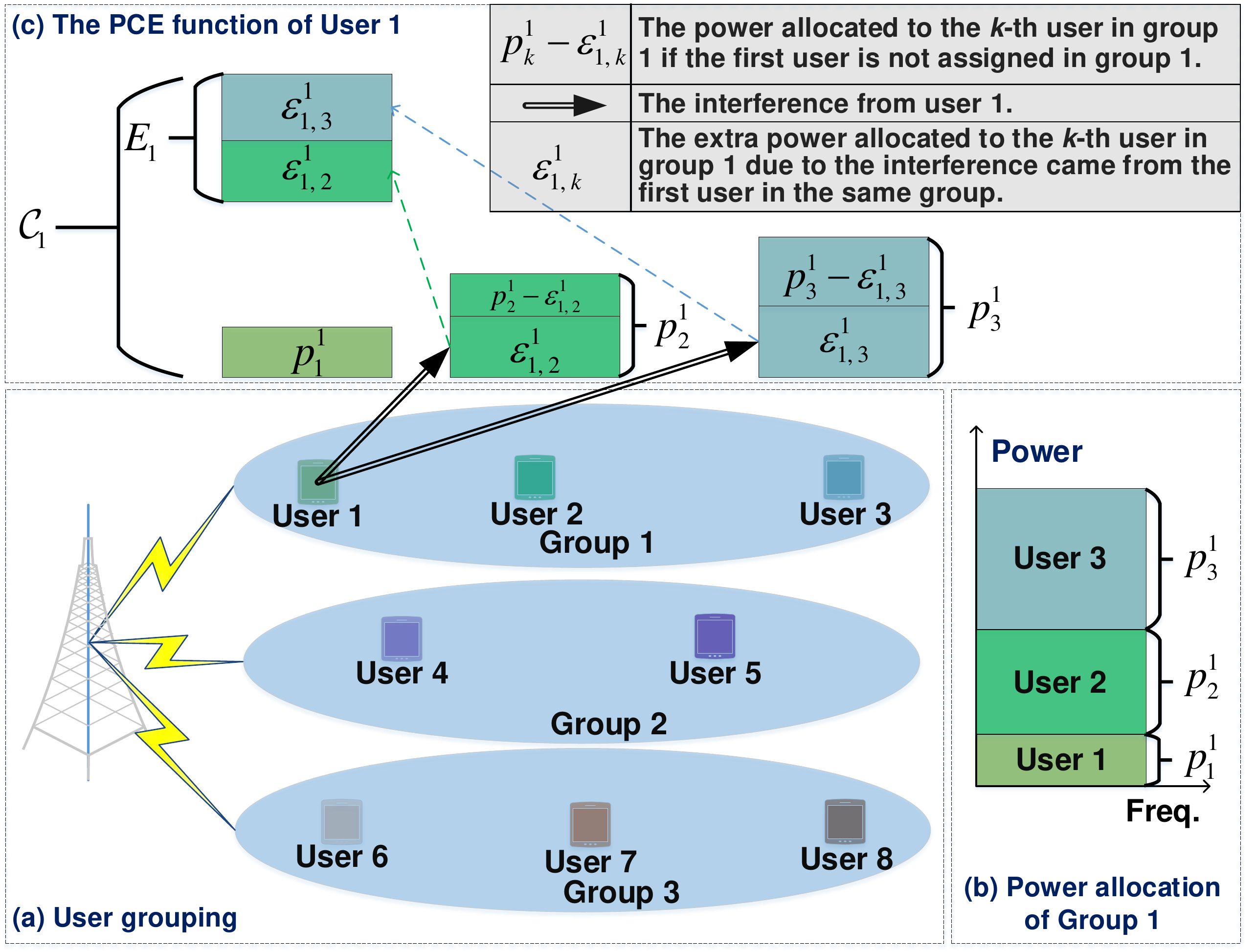}
	\caption{Illustration of User Grouping}
	\label{fig:0}
\end{figure}

A simple case of user grouping is illustrated in Fig. \ref{fig:0}(a), where eight users are assigned into three groups, and the distances between the BS and the users reflect the channel conditions of users. And as Fig. \ref{fig:0}(b) shows, more power is allocated to the user with worse channel condition according to the NOMA protocol \cite{Dai1}.


Let $\mathcal{U}^g$ denote the set of indexes of users assigned into group $g\in \mathcal{G}$, where the number of users in group $g$ is $\|\mathcal{U}^g\|$.  
 The transmit power allocated to user $n$ is denoted by $p_{n}$. Let $s_n$ denote the transmitted symbol of user $n$.
 According to the NOMA principle, the BS exploits the SC and broadcasts the signal $\sum_{n\in\mathcal{U}^g}\sqrt{p_{n}}s_n$ to all users in group $g$ as Fig. \ref{fig:0}(b) shows.
 The signal that user $n$ receives is given by
\begin{equation*}\label{2000}
y_n=h_{n}\sum_{i\in\mathcal{U}^g}\sqrt{p_{i}}s_i+\mathfrak{n}_{n},~~~~~n\in \mathcal{U}^g,
\end{equation*}where $\mathfrak{n}_{n}$ is Additive White Gaussian Noise (AWGN) with zero-mean and variance $\sigma^2$, i.e., $\mathfrak{n}_{n}\thicksim\mathcal{N}(0, \sigma^2)$. Based on SIC in NOMA \cite{Sai1}, at the receiver, the user with the poorest channel condition can detect its signal directly by treating the other users' signal as noise. On the other hand, the user with better channel condition can first detect and subtract the signals for the users with poorer channel conditions, and finally decode its own signal. In this way, for any user $n\in\mathcal{U}^g$, signals of the users with better channel conditions in group $g$ will be regarded as interference. Therefore, according to Shannon's theorem, the data rate achievable to user $n$ if user $n$ is assigned into group $g$ is given by \cite{Ding2014}:
\begin{equation}
\label{211}
R_{n} = log_2
\Bigg(1+\frac{|h_{n}|^2p_{n}}{|h_{n}|^2\!\!\!\!\!\!\!\sum\limits_{_{|h_{i}|>|h_{n}|}^{~~~\!i\in\mathcal{U}^g} }\!\!\!\!\!\!\!p_{i}+\sigma^2}
 \Bigg),~~~~~n\in \mathcal{U}^g.
\end{equation}

To meet the QoS requirements of all users, the achievable data rate of each user should be greater than the target data rate $r_n$, i.e.,
{\setlength\abovedisplayskip{1pt}
\setlength\belowdisplayskip{1pt}
\begin{equation}\label{212}
R_{n}\geq r_n,~~~~n\in \mathcal{U}^g.
\end{equation}}The total transmit power allocated to $N$ users is:
{\setlength\abovedisplayskip{1pt}
\setlength\belowdisplayskip{1pt}
\begin{equation}\label{213}
P_t=\sum\limits_{g\in \mathcal{G}}\sum\limits_{n\in \mathcal{U}^g}p_{n}.
\end{equation}}\section{Problem Formulation}
As described in Section II, OMA is carried out among different groups, there is no interference among users in different groups. According to ($\ref{211}$) and ($\ref{212}$), the power allocated to user $n\in \mathcal{U}^g$ should satisfy the following constraint:
{\setlength\abovedisplayskip{1pt}
\setlength\belowdisplayskip{1pt}
\begin{equation*}\label{3001-}
log_2
\Bigg(1+\frac{|h_{n}|^2p_{n}}{|h_{n}|^2\!\!\!\!\!\!\!\sum\limits_{_{|h_{i}|>|h_{n}|}^{~~i\in\mathcal{U}^g} }\!\!\!\!\!\!\!p_{i}+\sigma^2}
 \Bigg)\geq r_n,\quad n\in \mathcal{U}^g, g\in \mathcal{G},
\end{equation*}}where the equality holds with the minimal power consumption allocated to user $ n\in \mathcal{U}^g $:
{\setlength\abovedisplayskip{1pt}
\setlength\belowdisplayskip{1pt}
\begin{equation}\label{3001}
p_{n}=(2^{r_{n}}-1)(\frac{\sigma^2}{|h_{n}|^2}+\!\!\!\!\!\!\!\sum\limits_{_{|h_{i}|>|h_{n}|}^{~~i\in\mathcal{U}^g }}\!\!\!\!\!\!p_{i}) ,~~~~~n\in \mathcal{U}^g, g\in \mathcal{G}.
\end{equation}}

We can get the optimal power allocation strategy to minimize the total power for any given grouping strategies considering QoS requirements as stated in the Theorem \ref{thm1}:
\begin{thm1}
\label{thm1}
In order to minimize the total power consumption while guaranteeing the QoS requirements of all users, the optimal power allocation strategy for group $g\in\mathcal{G}$ is allocating power to each user according to ($\ref{3001}$) one by one from the user with the best channel condition to the user with the worst channel condition.
\end{thm1}
\begin{proof}
We w.l.o.g. assume that the channel conditions of the users in group $g$ have been ordered as $|h^g_{1}|^2\geq\cdots\geq|h^g_{\|\mathcal{U}^g\|}|^2$, where $h^g_{k}$ denotes the channel condition of the $k$-th user in group $g$. With SIC, the user with the best channel condition in each group would not be interfered by the other users. Therefore, the minimal power allocated to the user with the best channel condition in group $ g $, i.e. $ p_{1}^g $ is
{\setlength\abovedisplayskip{1pt}
\setlength\belowdisplayskip{1pt}
\begin{equation*}
\label{3002}
p^g_{1}=(2^{r^g_{1}}-1) \frac{\sigma^2}{|h^g_{1}|^2}.
\end{equation*}}where $p^g_{k}$ denotes the transmit power for the $k$-th user in group $g$, $r^g_{k}$ denotes the target data
rate of the $k$-th user in group $g$. Similarly, the minimal power allocated to the user with the second best channel condition in group $ g $ is
\begin{equation*}\label{3003}
p^g_{2}\!\!=\!\!(2^{r^g_{2}}\!-\!1)(\frac{\sigma^2\!\!}{|h^g_{2}|^2}\!+\!p^g_{1}) .
\end{equation*}
By repeating the process, we can get the best power allocation strategy of all users as stated in Theorem 1.\qedhere
\end{proof}

Theorem 1 holds for all possible combinations of user grouping strategies and can be used to obtain corresponding minimal transmit power.
Next what we need to do is to find a user grouping strategy, whose transmit power allocated by Theorem 1 is minimal. The problem of user grouping is formulated as follows:
{\setlength\abovedisplayskip{1pt}
\setlength\belowdisplayskip{1pt}
\begin{subequations}\label{3005}
	\begin{align}
	\min_{{\mathcal{U}}^g} ~~ &P_t=\sum\limits_{g\in\mathcal{G}}\sum\limits_{n\in\mathcal{U}^g}p_n\\
     \text{s.t.}~~~~&p_{n}=(2^{r_{n}}\!\!-\!1)\big(\frac{\sigma^2}{|h_{n}|^2}\!\!+\!\!\!\!\!\!\!\sum\limits_{_{|h_{i}|>|h_{n}|}^{~~i\in\mathcal{U}^g }}\!\!\!\!\!\!p_{i}\big),~~n\in \mathcal{U}^g, g\in \mathcal{G},\\
     &\bigcup\limits_{g\in \mathcal{G}}{\mathcal{U}^{g}}=\mathcal{N},\\
    &{\mathcal{U}^g}\cap{\mathcal{U}^{g'}}=\varnothing,~~~g,g'\in \mathcal{G},
	\end{align}
\end{subequations}}where ($\ref{3005}c$) and ($\ref{3005}d$) means that every user should be assigned into only one group.


The number of users in each group is arbitrary, hence the matching theory or the matching game used in existing strategies on user grouping cannot be applied. To solve the mixed-integer non-linear programming (MINLP) problem, in section IV, the PCE function is defined, and the relationship between the PCE function and the total power consumption is deduced.

\section{The PCE Function in Downlink NOMA}
In this section, we first analyze interference that a user brings to the others in the same group,
i.e., externalities in NOMA.

Let $ \pi_n $ denote the group index of user $ n $, i.e., $\pi_n\!=g\Leftrightarrow n\in \mathcal{U}^g$.
Let $\bm{\pi}$ denote the vector of the grouping strategies of all users,
i.e., $\bm{\pi}=(\pi_1,\cdots,\pi_N)$. Let $\bm{\pi}_{-n}$ denote the grouping strategies of all users in $\mathcal{N}\setminus\{n\}$,
i.e., $\bm{\pi}_{-n}=(\pi_1, \cdots, \pi_{n-1}, \pi_{n+1},\cdots, \pi_{N})$.

According to ($\ref{3001}$), the transmit power allocated to the user with the $k$-th best channel condition in group $g$ is
{\setlength\abovedisplayskip{1pt}
\setlength\belowdisplayskip{1pt}
\begin{equation}\label{4331}
 p^g_{k}=(2^{r^g_{k}}-1)\big(\frac{\sigma^2}{|h^g_{k}|^2}+\sum\limits_{{j=1}}^{k-1}p^g_{j}\big),~~~~k\leqslant \|\mathcal{U}^g\|.
 \end{equation}}Note that the transmit power allocated to the user with
the best channel condition in group $g$ is $p^g_{1}=(2^{r^g_{1}}\!-\!1)\frac{\sigma^2}{|h^g_{1}|^2}$. As shown in ($\ref{4331}$), the transmit power for a user will affect power allocated to the other users in the same group. Such external effects are called externalities. To exactly describe the externalities of the $k$-th user in group $g$ brings to the other users in group $g$, we define the extra power consumption that the $k$-th user in group $g$ brings to the $j$-th user in group $g$ as $\varepsilon^g_{k,j}$ ($k<j\leqslant \|\mathcal{U}^g\|$). According to ($\ref{4331}$), the transmit power allocated to the $(\!k\!+\!1\!)$-th user in group $g$ is
{\setlength\abovedisplayskip{1pt}
\setlength\belowdisplayskip{1pt}
\begin{equation}\label{4332}
\begin{aligned}
p^g_{k+1}\!=&(2^{r^g_{k+1}}\!\!-\!\!1)\big(\frac{\sigma^2}{|h^g_{k+1}|^2}\!+\!p^g_{k}\!+\!\!\sum\limits_{{j=1}}^{k-1}p^g_{j}\big)\\
=&(2^{r^g_{k+1}}\!\!-\!\!1)\big(\frac{\sigma^2}{|h^g_{k+1}\!|^2}\!\!+\!\!\sum\limits_{{j=1}}^{k-1}p^g_{j}\big)\!\!+\!p^g_{k}\!(2^{r^g_{k+1}}\!\!-\!\!1\!).
\end{aligned}
\end{equation}}Note that the transmit power allocated to the users with better channel conditions will not be affected by the $k$-th user in group $g$ as Theorem 1 states. Thus, the value of $\sum_{{j=1}}^{k-1}p^g_{j}$ in ($\ref{4332}$) will not be affected by the $k$-th user in group $g$. Accordingly, the power consumption that the $k$-th user in group $g$ brings to the $\!k\!+\!1$-th user in group $ g $ is
\begin{equation*}\label{4333}
\varepsilon^g_{k,k+1}=p^g_{k}(2^{r^g_{k+1}}-1).
\end{equation*}
According to ($\ref{4331}$) and ($\ref{4332}$), the transmit power allocated to the $(\!k\!+\!2\!)$-th user in group $g$ is
\begin{equation*}
\begin{aligned}
p^g_{k+2}\!\!=&(2^{r^g_{k+2}}\!\!-\!\!1)\big(\!\frac{\sigma^2}{|h^g_{k+2}\!|^2}\!+\!p^g_{k+1}\!\!+\!p^g_{k}\!\!+\!\!\sum\limits_{{j=1}}^{k-1}\!p^g_{j}\big)\\
=&(2^{r^g_{k+2}}\!\!-\!\!1\!)\!\Bigg(\!\frac{\sigma^2}{|h^g_{k+2}\!|^2}\!\!+\!\!(2^{r^g_{k+1}}\!\!-\!\!1\!)\!\big(\!\frac{\sigma^2}{|h^g_{k+1}\!|^2}\!\!+\!\!\sum\limits_{{j=1}}^{k-1}p^g_{j}\!\big)\\
&+\!\!\sum\limits_{{j=1}}^{k-1}p^g_{j}\!\!\Bigg)\!\!+\!\!(2^{r^g_{k+2}}\!-\!1)\!\Big(p^g_{k}(2^{r^g_{k+1}}\!\!-\!\!1)\!+\!p^g_{k}\!\Big),
\end{aligned}
\end{equation*}
where the power consumption that the $ k $-th user in group $ g $ brings to the $ (k\!+\!2\!) $-th user in group $g$ is
\begin{equation*}\label{4335}
\begin{aligned}
\varepsilon^g_{k,k+2}\!&=\!(2^{r^g_{\!k+2}}\!-\!1)\Big(p^g_{k}(2^{r^g_{\!k+1}}\!-\!1)\!+\!p^g_{k}\Big)\\
&=\!p^g_{k}2^{r^g_{k+1}}(2^{r^g_{k+2}}-1).
\end{aligned}
\end{equation*}
Similarly, we have
\begin{equation}\label{4000}\nonumber
\varepsilon^g_{k,j}\!=\!\left\{\begin{array}{ll}
\!\!\!p^g_{k}(2^{r^g_{j}}-1),\!\!\!\!\!\!\!\!\!& \textrm{$j\!=\!k\!+\!1$}\\
\!\!\!p^g_{k}\Big((2^{r^g_{j}}-1)\!\!\!\!\prod\limits_{_{k<i<j}}\!\!\!\!2^{r^g_{i}}\!\Big),~~~~\!\!\!\!\!\!\!\!\!& \textrm{$j\!>\! k\!+\!1$}\\
 \end{array} \right.\!\!.
\end{equation}
Hence, the total power consumption that the $ i $-th user in group $ g $ brings to the other users in group $ g $, i.e., the externalities function of the $ k $-th user in group $ g $, is
\begin{equation}
\label{4000+}
\begin{aligned}
E^g_{k}=&\sum\limits_{{j=i+1}}^{\|\mathcal{U}^g\|}\!\!\!\varepsilon^g_{k,j}\!\!
=p^g_{k}\!\!\!\prod\limits_{{j=i+1}}^{\|\mathcal{U}^g\|}\!\!\!2^{r^g_{j}}-p^g_{k},~~~g\in \mathcal{G}.
\end{aligned}
\end{equation}
Note that the externalities function of the user with the worst channel condition in group $ g $ is $ E^g_{\|\mathcal{U}^g\|}=0$.


Let $(n\in\mathcal{U}^g,\bm{\pi}_{-n})$ stand for the case that user $n$ is assigned into group $g$ and the grouping strategy of the other users is $\bm{\pi}_{-n}$. We define the Power Consumption and Externalities (PCE) function $\mathcal{C}_n$ according to ($\ref{4000+}$) as follows:
\begin{equation}\label{4001}
\begin{aligned}
\mathcal{C}_n(n\in\mathcal{U}^g,\bm{\pi}_{-n})=&p_{n}+E_{n}\\=&(2^{r_{n}}\!-\!1)\big(\!\!\!\!\sum\limits_{_{|h_{i}|>|h_{n}|}^{~~i\in\mathcal{U}^g }}\!\!\!\!\!\!p_{i}+\frac{\sigma^2}{|h_{n}|^2}\!\big)\!\!\!\!\!\!\prod\limits_{^{~~i\in \mathcal{U}^g}_{|h_{i}|<|h_{n}|}}\!\!\!\!\!\!2^{r_{i}}.
\end{aligned}
\end{equation}

According to the analysis above, $\mathcal{C}_n(n\in\mathcal{U}^g,\bm{\pi}_{-n})$ is also the total power consumption that user $n$ brings to group $g\in \mathcal{G}$, so we have
\begin{equation}\label{4001+}
\begin{aligned}
\!\!\!\!\!P^{g}(n\!\in\!\mathcal{U}^g,\bm{\pi}_{\!-n})\!-\!P^{g}(n\!\notin\!\mathcal{U}^g,\bm{\pi}_{\!-n})
\!=\!\mathcal{C}_n(n\in\mathcal{U}^g,\bm{\pi}_{-n}).
\end{aligned}
\end{equation}
where $P^{g}$ is the total transmit power allocated to the users in group $ g\in \mathcal{G}$. Accordingly, Proposition $\ref{Pro222}$ is obtained below.

\newtheorem{Pro1}{\bf Proposition}

\begin{Pro1}\label{Pro222}
	If we change the grouping strategy of one user, the difference of this user's PCE function $\mathcal{C}_n$ can reflect the difference of the total power consumption exactly.
\end{Pro1}

\begin{proof}
For $\forall n \in \mathcal{N}$, $\forall g,\tilde{g} \in \mathcal{G}$, and any given $\bm{\pi}_{-n}$, according to ($\ref{213}$) and ($\ref{4001+}$) we have
{\setlength\abovedisplayskip{1pt}
\setlength\belowdisplayskip{1pt}
\begin{equation*}
\label{4008}
\begin{aligned}
&P_t(n\in\mathcal{U}^g,\bm{\pi}_{-n})-P_t(n\in\mathcal{U}^{\tilde{g}},\bm{\pi}_{-n})	\\
&=P^g(n\in\mathcal{U}^g,\bm{\pi}_{-n}) - P^g(n\notin\mathcal{U}^g,\bm{\pi}_{-n})\\
&-\Big(P^{\tilde{g}}(n\in\mathcal{U}^{\tilde{g}},\bm{\pi}_{-n})-P^{\tilde{g}}(n\notin\mathcal{U}^{\tilde{g}},\bm{\pi}_{-n})\Big)\\
&=\mathcal{C}_n(n\in\mathcal{U}^g,\bm{\pi}_{-n})-\mathcal{C}_n(n\in\mathcal{U}^{\tilde{g}},\bm{\pi}_{-n}).
\end{aligned}
\end{equation*}}
Therefore, the proof of Proposition $\ref{Pro222}$ is concluded.\qedhere
\end{proof}

 A simple illustration of the PCE function is shown in Fig. \ref{fig:0}(c), where the transmit power for user $2$ and user $3$ is divided into two parts. One is the transmit power that user $1$ brings, i.e., $\varepsilon^1_{1,2}$ and $\varepsilon^1_{1,3}$, the other is the transmit power allocated to user $2$ and user $3$ if user $1$ is not in group $1$, i.e., $p^1_2-\varepsilon^1_{1,2}$ and $p^1_3-\varepsilon^1_{1,3}$. Then the PCE function of user $1$ is $p^1_1+\varepsilon^1_{1,2}+\varepsilon^1_{1,3}$, which is the total power consumption that user $1$ brings to group $1$.

\section{Directed Graph for User Grouping in Downlink NOMA}

In this section, based on the definition of PCE, we first build a directed graph. Then, we attempt to solve the user grouping problem with QoS constrains by graph theory. In order to extend Proposition 1 to multi-user scenarios, we introduce the following definitions.

\begin{def1}\label{defe1}
For any $i$ users numbered $\{n_1,\cdots ,n_i\}$ in different groups,
if we change the grouping strategy $\bm{\pi}$ into $\tilde{\bm{\pi}}$ by $(n_1\!\!\!\!\rightarrow\!\!\!\!\pi_{n_2}),\cdots ,(n_{i-1}\!\!\!\rightarrow \!\!\pi_{n_i})$, and the total power consumption satisfies $ P_{t}(\tilde{\bm{\pi}}) < P_{t}(\bm{\pi}) $, then these $ i $ users compose an \emph{i-shift league}.
\end{def1}

Where $(n\!\!\rightarrow\!\! g)$ stand the operation that user $n$ is moved into group $g$. Similarly, $(n\!\!\rightarrow\!\! \pi_{\tilde{n}})$ stand the operation that user $n$ is moved into the group including user $\tilde{n}$.

\begin{def1}\label{defe2}
For any $i$ users numbered $\{n_1,\cdots ,n_i\}$ in different groups. If we change the grouping strategy $\bm{\pi}$ into $\tilde{\bm{\pi}}$ by $(n_1\rightarrow \pi_{n_2}),\cdots,(n_{i-1}\rightarrow \pi_{n_i}),(n_i\rightarrow \pi_{n_1})$, and the total power consumption satisfies $ P_{t}(\tilde{\bm{\pi}})<P_{t}(\bm{\pi}) $, then these $ i $ users compose an \emph{i-exchange league}.
\end{def1}
\begin{def1}\label{defe3}
A grouping strategy $\bm{\pi}$ is called \emph{all-stable solution}, if for $ \forall i \in \mathcal{N} $, there is no \emph{i-shift league} or \emph{i-exchange league} in them.
\end{def1}

However, to find the \emph{all-stable solution} of the problem in (\ref{3005}) is an obstacle. We will resort to graph theory.

\subsection{Directed Graph}

To apply graph theory more conveniently, we create a virtual user for every group, and let $ \mathcal{N}^v = \{N+1, \cdots, N+G\} $ denote the set of indexes of virtual users. User $(N+g)$ is the virtual user in group $g\in\mathcal{G}$, i.e. $\pi_{N+1}=1,\cdots,\pi_{N+G}=G$.
Let $ \bm{\pi}^e $ denote the vector of the grouping strategies of all users in $\mathcal{N}^e=\mathcal{N}\cup\mathcal{N}^v$. 
These virtual users are used to find the \emph{shift league}s to be stated later.

Let the target data rates of virtual users be 0. Then according to ($ \ref{3001} $) and ($ \ref{4001} $), the power consumption and the PCE function of the virtual users are 0, i.e.,
\begin{equation}\label{3005e+11}
p_{n}=\mathcal{C}_{n}({n}\in\mathcal{U}^{g},\bm{\pi}_{-{n}}^e)=0,~~~~g\in\mathcal{G},n\in\mathcal{N}^v.
\end{equation}
Therefore, the transmit power allocated to the real users will not be affected by the virtual users.

Let $(n\twoheadrightarrow \tilde{n})$ stand the operation that user $n$ is moved into the group including user $\tilde{n}$, and user $\tilde{n}$ is moved out of group $\pi_{\tilde{n}}$. Assume that $\pi_n=g$ and $\pi_{\tilde{n}}=\tilde{g}$. Then the difference of the PCE function of user $n$ before and after $(n\twoheadrightarrow \tilde{n})$ is
\begin{equation*}\label{3005e+3}
\begin{aligned}
&\bigtriangleup_{n}(n\twoheadrightarrow \tilde{n},\bm{\pi}^e_{-\{n,\tilde{n}\}})\\
=&\mathcal{C}_{n}({n}\in\mathcal{U}^{\tilde{g}},{\tilde{n}}\not\in\mathcal{U}^{\tilde{g}},\bm{\pi}_{-\{n,\tilde{n}\}}^e)-\mathcal{C}_{n}({n}\in\mathcal{U}^g,\bm{\pi}_{-{n}}^e)\\
=&(2^{r_{n}}\!\!-\!1)\Big(\!\!\big(\!\!\!\!\!\!\!\sum\limits_{_{~~\!|h_{i}|>|h_{n}|}^{~~\!i\in\mathcal{U}^{\tilde{g}}\setminus \{\tilde{n}\} }}\!\!\!\!\!\!p_{i}+\frac{\sigma^2}{|h_{n}|^2}\!\big)\!\!\!\!\!\!\!\!\prod\limits_{^{~\!\!i\in \mathcal{U}^{\tilde{g}}\setminus \{\tilde{n}\}}_{|h_{i}|<|h_{{n}}|}}\!\!\!\!\!\!2^{r_{i}}\!\!-\!\!\big(\!\!\!\!\!\sum\limits_{_{|h_{i}|>|h_{n}|}^{~~i\in\mathcal{U}^g }}\!\!\!\!\!\!p_{i}+\frac{\sigma^2}{|h_{n}|^2}\!\big)\!\!\!\!\!\!\!\!\prod\limits_{^{~~i\in \mathcal{U}^g}_{|h_{i}|<|h_{{n}}|}}\!\!\!\!\!\!2^{r_{i}}\!\!\Big),
\end{aligned}
\end{equation*}

A directed graph $G(\mathcal{V},\mathcal{E};\bm{\pi}^e)$ is constructed, where $\mathcal{V}\!\!=\!\!\mathcal{N}^e$ is the set of nodes, $\mathcal{E}$ is the set of edges and the edges only exist between two users in different groups. The adjacent matrix $\mathbf{A}=(a_{ij})_{i,j\in \mathcal{N}}$ is formed as follows:
\begin{equation}\label{3005e+12}
a_{ij}=\left\{\begin{array}{ll}\!\!\!
\bigtriangleup_{i}(i\twoheadrightarrow j,\bm{\pi}^e_{-\{i,j\}}),
 \!\!\!\!\!~~~~~~& \pi_i\neq \pi_j\\
\!\!\!\infty,
 & else\\
 \end{array} \right..
\end{equation}
where $i,j\in\mathcal{V}$ and $a_{ij}=\infty$ means that node $i$ is not connected with node $j$. Then the following lemma about this graph can be obtained.

%
%
%
%

\begin{lem1}\label{lem1}
For any $i$ users in different groups, the following propositions:

$\textcircled{\small{\small{1}}}$ These $i$ users can compose an \emph{i-exchange league}.

$\textcircled{\small{\small{2}}}$ These $i$ users can compose a negative loop in graph $G(\mathcal{V},\mathcal{E};\bm{\pi}^e)$.

are equivalent.

\end{lem1}

\begin{proof}

Assume that in grouping strategy $\bm{\pi}^e$, users $n_1,\cdots,n_i$ are assigned into different groups, and $\pi_{n_1}=g_1,\cdots,\pi_{n_i}=g_i$. If we change the grouping strategy $\bm{\pi}^e$ by $(n_1\!\!\rightarrow\!\! g_2),\cdots,(n_{i-1}\!\rightarrow \!\! g_i),(n_i\!\!\rightarrow\!\! g_1)$, according to ($\ref{4001+}$), the difference of the power consumption of group $g_k(k\leqslant i)$ is
\begin{equation}\label{3005e+5}
\begin{aligned}
&\bigtriangleup^{g_k}(n_{k'}\twoheadrightarrow n_k,\bm{\pi}_{-\{n_{k'},n_k\}}^e)\\
=&\mathcal{C}_{n_{{k'}}}\!({n_{{k'}}}\!\in\!\mathcal{U}^{g_k}\!,\!{n_k}\!\not\in\!\mathcal{U}^{g_k}\!,\!\bm{\pi}_{\!-\{n_{{k'}},n_k\}}^e\!)\!-\!\mathcal{C}_{n_k}\!({n_k}\!\!\in\mathcal{U}^{g_k}\!,\bm{\pi}_{-{n_k}}^e\!)\\
\end{aligned}
\end{equation}
where ${k'}\!=(\!(k\!-\!2)\!\!\!\!\mod i)\!+\!1$. So $\textcircled{\small{\small{1}}}$ is equivalent to that the difference of total power consumption is negative, i.e.,
{\setlength\abovedisplayskip{1pt}
\setlength\belowdisplayskip{1pt}
\begin{equation}\label{3005e+6}
\begin{aligned}
&\bigtriangleup_{P_{t}}\Big((n_1\rightarrow g_2),\cdots,(n_{i-1}\rightarrow g_i),(n_i\rightarrow g_1)\Big)\\
=&\sum\limits_{k=1}^{i}\bigtriangleup^{g_k}(n_{{k'}}\twoheadrightarrow n_k,\bm{\pi}_{-\{n_{{k'}},n_k\}}^e)\\
=&\sum\limits_{k=1}^{i}\mathcal{C}_{n_{{k'}}}\!({n_{{k'}}}\!\in\!\mathcal{U}^{g_k}\!\!,{n_k}\!\not\in\!\mathcal{U}^{g_k}\!,\bm{\pi}_{\!-\{n_{{k'}},n_k\}}^e\!)\!\\&-
\!\sum\limits_{k=1}^{i}\mathcal{C}_{n_k}\!({n_k}\in\mathcal{U}^{g_k}\!,\bm{\pi}_{-{n_k}}^e\!)<0\\
\end{aligned}
\end{equation}}

$\textcircled{\small{\small{2}}}$ is equivalent to that the total edge weight of loop $n_1-\!\!>\!n_2-\!\!>\cdots-\!\!>\!n_{i-1}-\!\!>\!n_i-\!\!>\!n_1$ is negative, i.e.,
{\setlength\abovedisplayskip{1pt}
\setlength\belowdisplayskip{1pt}
\begin{equation}\label{3005e+8}
\begin{aligned}
\sum\limits_{k=1}^{i}a_{{k'}k}=&\sum\limits_{k=1}^{i}\bigtriangleup_{n_{{k'}}}(n_{{k'}}\twoheadrightarrow n_k,\bm{\pi}_{-\{{k'},k\}}^e)\\
=&\sum\limits_{k=1}^{i}\mathcal{C}_{n_{{k'}}}({n_{{k'}}}\in\mathcal{U}^{g_{{k}}},{n_{k}}\not\in\mathcal{U}^{g_{{k}}},\bm{\pi}_{-\{n_{{k}},n_{{k'}}\}}^e)\\
&-\sum\limits_{k=1}^{i}\mathcal{C}_{n_{{k'}}}({n_{{k'}}}\in\mathcal{U}^{g_{{k'}}},\bm{\pi}_{-{n_{{k'}}}}^e)<0\\
\end{aligned}
\end{equation}}
Obviously,
{\setlength\abovedisplayskip{1pt}
\setlength\belowdisplayskip{1pt}
\begin{equation}\label{3005e+9}
\sum\limits_{k=1}^{i}\begin{aligned}\mathcal{C}_{n_k}\!({n_k}\in\mathcal{U}^{g_k}\!,\bm{\pi}_{-{n_k}}^e\!)=\sum\limits_{k=1}^{i}\mathcal{C}_{n_{{k'}}}({n_{{k'}}}\in\mathcal{U}^{g_{{k'}}},\bm{\pi}_{-{n_{{k'}}}}^e)
\end{aligned}
\end{equation}}
Therefore, according to ($\ref{3005e+6}$), ($\ref{3005e+8}$) and ($\ref{3005e+9}$), the proof of Lemma $\ref{lem1}$ is concluded.\qedhere
\end{proof}

The transmit power can be reduced by finding the \emph{exchange league}s, and updating the grouping strategy. However, the user number in each group will remains unchanged. To find the better grouping strategy, the user number in each group should not be fixed. Accordingly, we need to find the \emph{shift league}s to improve the grouping strategy with arbitrary user numbers in each group.

\begin{lem1}
\label{Lemma2}
All \emph{shift league}s can be converted the into \emph{exchange league}s.
\end{lem1}

\begin{proof}
Assume that in grouping strategy $ \bm{\pi}^e $, users $ n_1, \cdots, n_i $ can compose an \emph{i-shift league}, i.e., the total power consumption can be reduced by $(n_1\rightarrow \pi_{n_2}),\cdots,(n_{i-1}\rightarrow \pi_{n_i})$. Assume that user $n_v$ is the virtual user in group $\pi_{{n_i}}$. If we change the grouping strategy $\bm{\pi}^e$ by $(n_1\rightarrow \pi_{n_2}),\cdots,(n_{i-1}\rightarrow \pi_{n_v}),(n_v\rightarrow \pi_{n_1})$, the total power consumption will be reduced as the power consumption of the real users will not be
affected by the virtual users. Therefore, users $ n_1, \cdots, n_{i-1}, n_v$ can compose an \emph{i-exchange league}.\qedhere
\end{proof}

For the convenience of description, the negative loop with all users in different groups is called the \emph{negative differ-group loop}. According to Lemma 1 and Lemma 2, Theorem $\ref{thm+1}$ and Theorem $\ref{thm+111}$ can be obtained.

\begin{thm1}\label{thm+1}
If there is no \emph{negative differ-group loop} in the directed graph $G(\mathcal{V},\mathcal{E};\bm{\pi}^e)$, the grouping strategy of real users $\bm{\pi}$ is \emph{all-stable solution}.
\end{thm1}

\begin{proof}

If $ i $ users can compose an \emph{i-exchange league}, according to Lemma 1, there must be a \emph{negative differ-group loop} in graph $G(\mathcal{V},\mathcal{E};\bm{\pi}^e)$. In addition, if $ i $ users can compose an \emph{i-shift league}, according to Lemma 2, there must be an \emph{i-exchange league} in $\mathcal{N}^e$. So there must be a \emph{negative differ-group loop} in graph $G(\mathcal{V},\mathcal{E};\bm{\pi}^e)$. Based on above analysis, if there is no \emph{negative differ-group loop} in graph $G(\mathcal{V},\mathcal{E};\bm{\pi}^e)$, there is no \emph{i-shift league} or \emph{i-exchange league} in $\bm{\pi}^e$ and $\bm{\pi}$ for $ \forall i \in \mathcal{G} $. So the proof of Theorem $ \ref{thm+1} $ is concluded.
\qedhere
\end{proof}

\begin{thm1}\label{thm+111}

If there is a \emph{negative differ-group loop} $n_1\!-\!>\!\cdots\!-\!\!>\!n_i-\!\!>\!n_1\!$ in graph $G(\mathcal{V},\mathcal{E};\bm{\pi}^e)$, the total power allocated to the real users can be reduced by $(n_1\!\rightarrow\! \pi_{n_2}),\cdots,(n_{i-1}\!\rightarrow\! \pi_{n_i}),(n_i\!\rightarrow\! \pi_{n_1})$.
\end{thm1}

\begin{proof}

%

If $n_1-\!\!>\!\cdots\!-\!\!>\!n_i-\!\!>\!n_1$ is a \emph{negative differ-group loop}, according to ($\ref{3005e+6}$),($\ref{3005e+8}$) and ($\ref{3005e+9}$), we have
\begin{equation*}\label{3005e+21}
\begin{aligned}
\sum\limits_{k=1}^{i}\!a_{{k'}k} 
\!=\!\bigtriangleup_{P_{t}}\!\Big(\!\!(n_1\!\!\rightarrow\!\! \pi_{n_2}\!),\!\cdots\!,\!(n_{i-1}\!\!\rightarrow\!\! \pi_{n_i}\!),(n_i\!\!\rightarrow\!\! \pi_{n_1}\!)\!\!\Big)\!\!<\!\!0.
\end{aligned}
\end{equation*}
Therefore, the total transmit power allocated for all users in $ \mathcal{N}^e $ can be reduced by $ (n_1\rightarrow \pi_{n_2}),\cdots,(n_{i-1}\rightarrow \pi_{n_i}),(n_i\rightarrow \pi_{n_1}) $. As mentioned in ($ \ref{3005e+11}$), the transmit power allocated to virtual users is 0, hence the total transmit power allocated to the real users can also be reduced. So the proof of Theorem $ \ref{thm+111} $ is concluded. \qedhere
\end{proof}
{\setlength\abovedisplayskip{1pt}
\setlength\belowdisplayskip{1pt}
\begin{figure}[ht]
	\centering
	\includegraphics[scale=0.35]{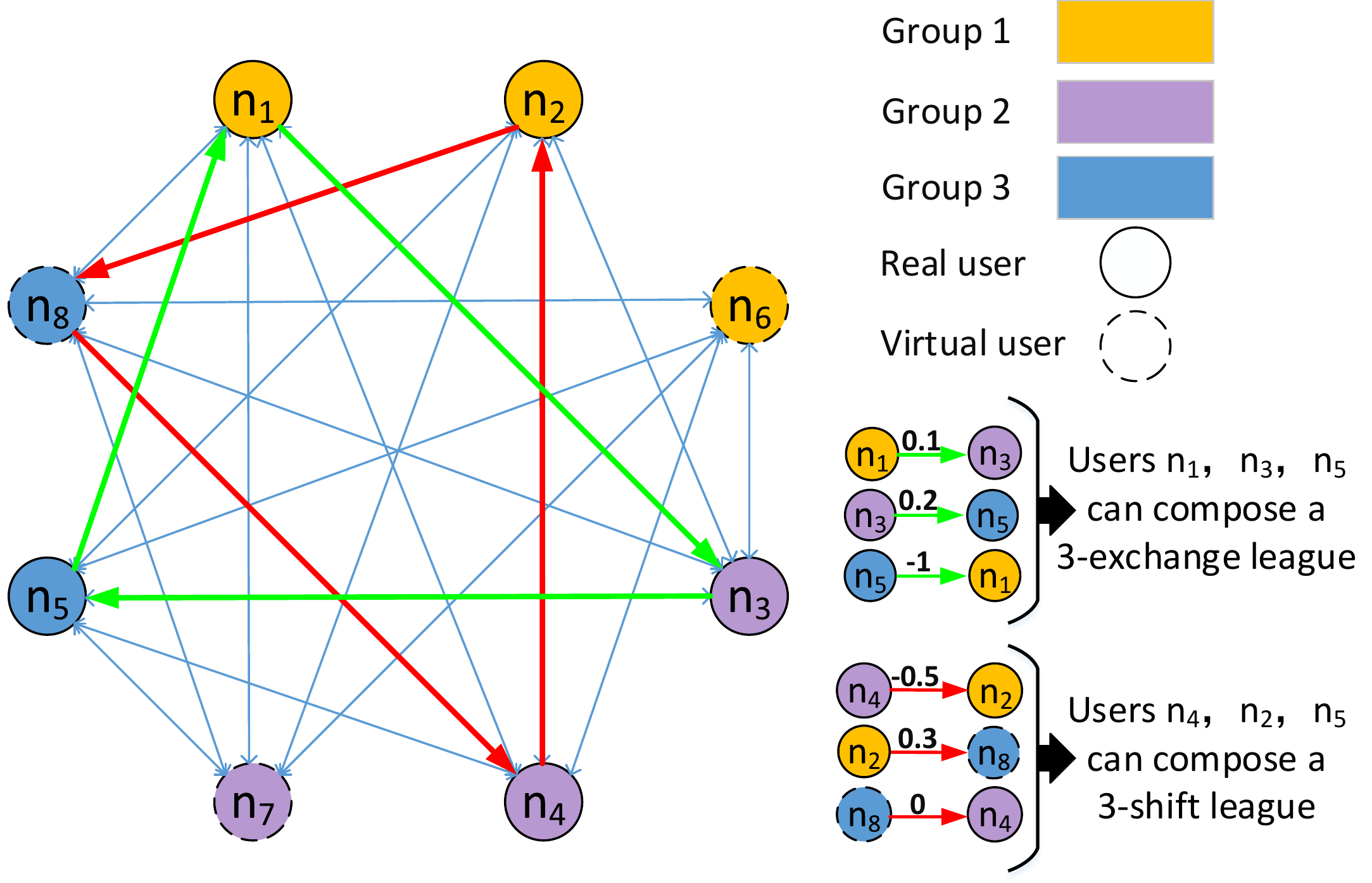}
	\caption{Illustration of the Directed Graph}
	\label{fig:7}
\end{figure}}
A simple case of the directed graph is illustrated in Fig. \ref{fig:7}, where five users(with solid borders) are assigned into three groups, and the virtual users (with dashed borders) are added into each group. The edges in this graph only exist between the users in different groups. $n_1-\!\!>\!n_3\!-\!\!>\!n_5-\!\!>\!n_1$ is a \emph{negative differ-group loop}, and $n_1$, $n_3$, $n_5$ are real users, so users $n_1,n_3,n_5$ can compose a 3-exchange league. In this case, we can reduce the total power consumption by $(n_1\rightarrow \pi_{n_3}),(n_{3}\rightarrow \pi_{n_5}),(n_5\rightarrow \pi_{n_1})$. Meanwhile, $n_4-\!\!>\!n_2\!-\!\!>\!n_8-\!\!>\!n_4$ is a \emph{negative differ-group loop}, and user $n_8$ is a virtual user, so users $n_4,n_2,n_5$ can compose a 3-shift league ($n_8,n_5\in \mathcal{U}^3$). That means we can reduce the total power consumption by $(n_4\rightarrow \pi_{n_2}),(n_{2}\rightarrow \pi_{n_8}),(n_{8}\rightarrow \pi_{n_4})$.


\subsection{Algorithm Description}
According to Theorem $\ref{thm+1}$ and Theorem $\ref{thm+111}$, the total power consumption can be reduced by searching for the \emph{negative differ-group loop}s in graph $G(\mathcal{V},\mathcal{E};\bm{\pi}^e)$, and changing the grouping strategy as Theorem $\ref{thm+111}$ states, until there is no \emph{negative differ-group loop} in graph $G(\mathcal{V},\mathcal{E};\bm{\pi}^e)$. The detailed main algorithm is described in Algorithm 1, where the user grouping is initialized in step 1-3 by grouping every user into the group with the best channel condition. Step 4-11 in Algorithm 1 is an iterative process to converge to all-stable solution.

\begin{algorithm}
 \caption{Graph Theory Based User Grouping Algorithm}\label{alg:1}
  \KwIn{Set of users $\mathcal{N}$, Set of groups $\mathcal{G}$,
  	 Target data rate $ {r}_n$, Channel condition $ {h}_{n}$, ${n\in\mathcal{N}}$}
  \KwOut{Power ${p}_n, n\in\mathcal{N}$, Grouping $\mathcal{U}$}
  \For{$n=1 \textrm{ to }N$}{
   Find $g_{min}=arg\min\limits_{^{g\in\mathcal{G}}_{n\in\mathcal{U}^g}}\big(|h_{n}|_{n\in\mathcal{U}^g}\big)$;
   $\mathcal{U}^{g_{min}}\leftarrow\mathcal{U}^{g_{min}}\cup \{n\}$;}
   \Repeat{$\mathcal{U}$ and $\bm{\pi}^e$ don't change}{
         Create graph $G(\mathcal{V},\mathcal{E};\bm{\pi}^e)$;

         Calculate the adjacent matrix $\mathbf{A}=(a_{ij})_{i,j\in\mathcal{V}^e}$ of graph $G(\mathcal{V},\mathcal{E};\bm{\pi}^e)$ according to ($\ref{3005e+12}$);

         Find the \emph{negative differ-group loop}s in graph $G(\mathcal{V},\mathcal{E};\bm{\pi}^e)$;

        \If{Find a \emph{negative differ-group loop} $\mathcal{L}$}{
		 Update $\mathcal{U}$ and $\bm{\pi}^e$ according to Theorem $\ref{thm+111}$;
}
}

   \For{$n=1 \textrm{ to }N$}{
         Calculate $p_{n}$ according to ($\ref{3001}$).
		}
		 \textbf{return} ${p}_n, n\in\mathcal{N}$, $\mathcal{U}$

\end{algorithm}
\begin{cor1}\label{Cor1}
Algorithm 1 can converge to \emph{all-stable solution} in finite iterations.
\end{cor1}

\begin{proof}
As $ \mathcal{G} $ and $ \mathcal{N} $ are both finite, the strategic space is finite. Furthermore, every time we change a user's grouping strategy, the total power consumption will be reduced. Based on these facts, Algorithm 1 is bound to stop at \emph{all-stable solution} in finite steps. So the proof of Corollary 1 is concluded.\qedhere
\end{proof}

Bellman-Ford algorithm is extended to find the \emph{negative differ-group loop}s in graph $G(\mathcal{V},\mathcal{E};\bm{\pi}^e)$\cite{fakcharoenphol2006planar} in step 7 of Algorithm 1.
We add a super node with outgoing edges to all the nodes in $\mathcal{V}$, and the main procedure of BellmanFord algorithm is searching for the shortest path from the super node to all other nodes by relaxation until none path can be relaxed\cite{bellman1958routing, ford2015flows}. Different from the original Bellman-Ford algorithm, during the relaxation steps, it is avoided that two different users in the same group appear in the same path. Therefore, if the number of nodes in the path from the super node to a node is greater than $G$, there must be a \emph{negative differ-group loop}s in this path. The detailed procedure of the extended Bellman-Ford algorithm is described in Algorithm 2.
In Algorithm 2, the path from super node to node $n$, i.e., $\mathcal{T}_n$, $n\in \mathcal{V}$, is relaxed by step 6 or step 11 repeatedly until all $\mathcal{T}_n$ do not change. Step 10 is a recursive procedures by calling Algorithm 2 repeatedly, its computational complexity increases exponentially with the dimensions of $N$ and $G$.

\begin{algorithm}
 \caption{Extended BellmanFord Algorithm for Finding the \emph{Negative Differ-Group Loop}}\label{alg:2}
  \KwIn{Group $G(\mathcal{V},\mathcal{E};\bm{\pi}^e)$,

  \quad \quad \quad Adjacent matrix$\mathbf{A}=(a_{ij})_{i,j\in\mathcal{V}}$}
  \KwOut{\emph{Negative differ-group loop} $\mathcal{L}$}

  Let the distance from super node to node $n$ be $\!m_n\!\!=\!\!0$, the path from super node to node $n$ be $\mathcal{T}_n=\varnothing$, $n\in \mathcal{V}$;

  \Repeat{$\mathcal{T}_n$ $ n\in\mathcal{V}$ do not change}{
        \For{$i=1 \textrm{ to }\|\mathcal{V}\|$}{
        \For{$j=1 \textrm{ to }\|\mathcal{V}\|$}{
        \If{$m_{j}\!>\!m_{i}+\!a_{ij}$ \& $ a_{kj}\neq \infty, \forall k\in\mathcal{T}_i\setminus \{j\}$}{
         $\mathcal{T}_j\leftarrow\mathcal{T}_i\cup \{i\}$, $m_{j}\leftarrow m_{i}+a_{ij}$
        }
        \If{$m_{j}\!>\!m_{i}+\!a_{ij}$ \& $\exists a_{kj}= \infty,k\in\mathcal{T}_i\setminus \{j\}$}{
         Find the shortest path $\mathcal{T}'_i$ from the super node to user $i$ with $ a_{kj}\neq \infty, \forall k\in\mathcal{T}'_i\setminus \{j\}$, assume the distance of path $\mathcal{T}'_i$ is $m'_{i}$;

        \If{$m_{j}>m'_{i}+a_{ij}$}{
         $\mathcal{T}_j\leftarrow\mathcal{T}'_i\cup \{i\}$, $m_{j}\leftarrow m'_{i}+a_{ij}$
		}
        }
        \If{$\|\mathcal{T}_j\|> G$}{
         Find the \emph{negative differ-group loop} $\mathcal{L}$ in $\mathcal{T}_j$;
         \textbf{return} $\mathcal{L}$ and break;
		}

		}

		}
          }

\end{algorithm}


To approach the solution of the problem in ($\ref{3005}$) with polynomial time computational complexity, we design a suboptimal fast algorithm based on a greedy strategy to find the \emph{negative differ-group loop}s in step $7$ of Algorithm $1$.
First we search for the minimum edge $a_{n_1n_2}$ in $\mathbf{A}$, and calculate the total edge weight of loop $n_1-\!\!>\!n_2-\!\!>\!n_1$,
next we search for the minimum output edge $a_{n_2n_3}$($\pi_{n_3}\neq\pi_{n_1},\pi_{n_3}\neq\pi_{n_2}$) of node $n_2$, and calculate the total edge weight of loop $n_1-\!\!>\!n_2-\!\!>\!n_3-\!\!>\!n_1$.
 Then we search for the next output edge as the same operation until the loop $n_1-\!\!>\!\cdots-\!\!>\!n_G-\!\!>\!n_1$, and set $a_{n_1n_2}=\infty$.
 The number of steps above repeat more than the number of edges, i.e., $(G+N)^2$. Then the loop with the minimum total edge weight is outputted if this loop is negative. To reduce unnecessary computation, the steps above are repeated $\lceil\alpha(N+G)\rceil$ times ($\alpha\in [\frac{1}{N+G},N]$), where $\alpha$ is a regulating factor to limit the number of iterations. The detailed procedure of the fast greedy  algorithm is described in Algorithm 3. In order to limit the iterations of Algorithm 3, it is avoid that a user be selected more than once as the first users in the \emph{negative differ-group loop}.

\begin{algorithm}
 \caption{Fast Greedy Algorithm for Finding the \emph{Negative Differ-Group Loop}}\label{alg:1}
  \KwIn{Group $ G(\mathcal{V},\mathcal{E};\bm{\pi}^e)$,

  \quad \quad \quad Adjacent matrix$\mathbf{A}=(a_{ij})_{i,j\in\mathcal{V}}$}
  \KwOut{\emph{Negative differ-group loop} $\mathcal{L}$}
  Let $\mathcal{T}$ record the path of each iteration, and $m_t$ record the total edge weights of this path;

   Let $m_s$ be the total edge weights of \emph{negative differ-group loop} $\mathcal{L}$;$m_s\leftarrow 0$;

        \For{$x=1 \textrm{ to }\lceil\alpha(N+G)\rceil$}{
        $\mathcal{T}=\varnothing$;

         Find the minimal edge $a_{ij}=\min\limits\{a_{ij}|i,j\in\mathcal{V}\}$;

        $\mathcal{T}\leftarrow\mathcal{T}\cup \{i\}\cup \{j\}$, $m_t\leftarrow a_{ij}+a_{ji}$;

        $a_{ij}\leftarrow\infty$, $i\leftarrow j$;

        \For{$l=3 \textrm{ to }G$}{
        Find the minimal output edge of node $i$ : $a_{ij}=\min\limits\{a_{ij}|j\in\mathcal{V}, a_{jk}\neq\infty, \forall k\in \mathcal{T}\}$;

        $m_t\leftarrow m_t+a_{ij}$, $\mathcal{T}\leftarrow\mathcal{T}\cup \{j\}$;

        \If{$m_s>m_t+a_{ji}$}{
        $\mathcal{L}\leftarrow\mathcal{T}$, $m_s\leftarrow m_t+a_{ji}$;
		}
        $i\leftarrow j$;
        }
        }
        \textbf{return} $\mathcal{L}$.

\end{algorithm}

\emph{Computational Complexity Analysis:} In Algorithm 3, steps $ 8 $-$ 15 $ require complexity of $O\big(G(G+N)\big)$. Hence, computational complexity of Algorithm 3 is $O\big( G(G+N)^2\big)$. Assume that Algorithm 3 is repeated $C$ times in Algorithm 1, then we have $C\leqslant N+G$ for a user would not be selected to be the first user in the negative differ-loop. Therefore, computational complexity of the proposed fast user grouping algorithm is $O\big(G(G+N)^3\big)$. In fact, $C$ is much less than $(G+N)$ as the simulation shows in section VI.

\section{Performance Evaluation}

In this section, we compare the performance of the proposed algorithms with two existing user grouping strategies in downlink NOMA. In the simulations, the BS is located in the cell center and the users are randomly distributed in a circular range with a radius of 500m. The minimum distance from users to BS is set to 35m.
The large-scale channel gain is $128.1\!+\!37.6$log$_{10}( d_{n}[$km$])$ dB.
The Rayleigh fading coefficient follows an i.i.d. Gaussian distribution as $\beta\!\!\thicksim\!\mathcal{CN}(0, 1)$. The noise power is $\sigma^2=BN_0$, where the bandwidth of each channel is $B\!\!= \!\!180$ kHz and the noise power spectral density is $N_0\!= $-$174$ dBm/Hz. The targeted data rate of each user is randomly chosen from 0.5 to 8 bps/Hz. The number of users is varied from $25$ to $300$ in our simulations.


\subsection{Impacts of $ \alpha $}

We will first evaluate the total power consumption in the extended BellmanFord algorithm(``Proposed (Bellman-Ford)''), and the proposed fast algorithm based on the greedy strategy with different $ \alpha $(``Proposed (Greedy)'').

\begin{figure}
	\centering
	\subfigure[]{
		\label{fig:6}
		\includegraphics[width=1.59in]{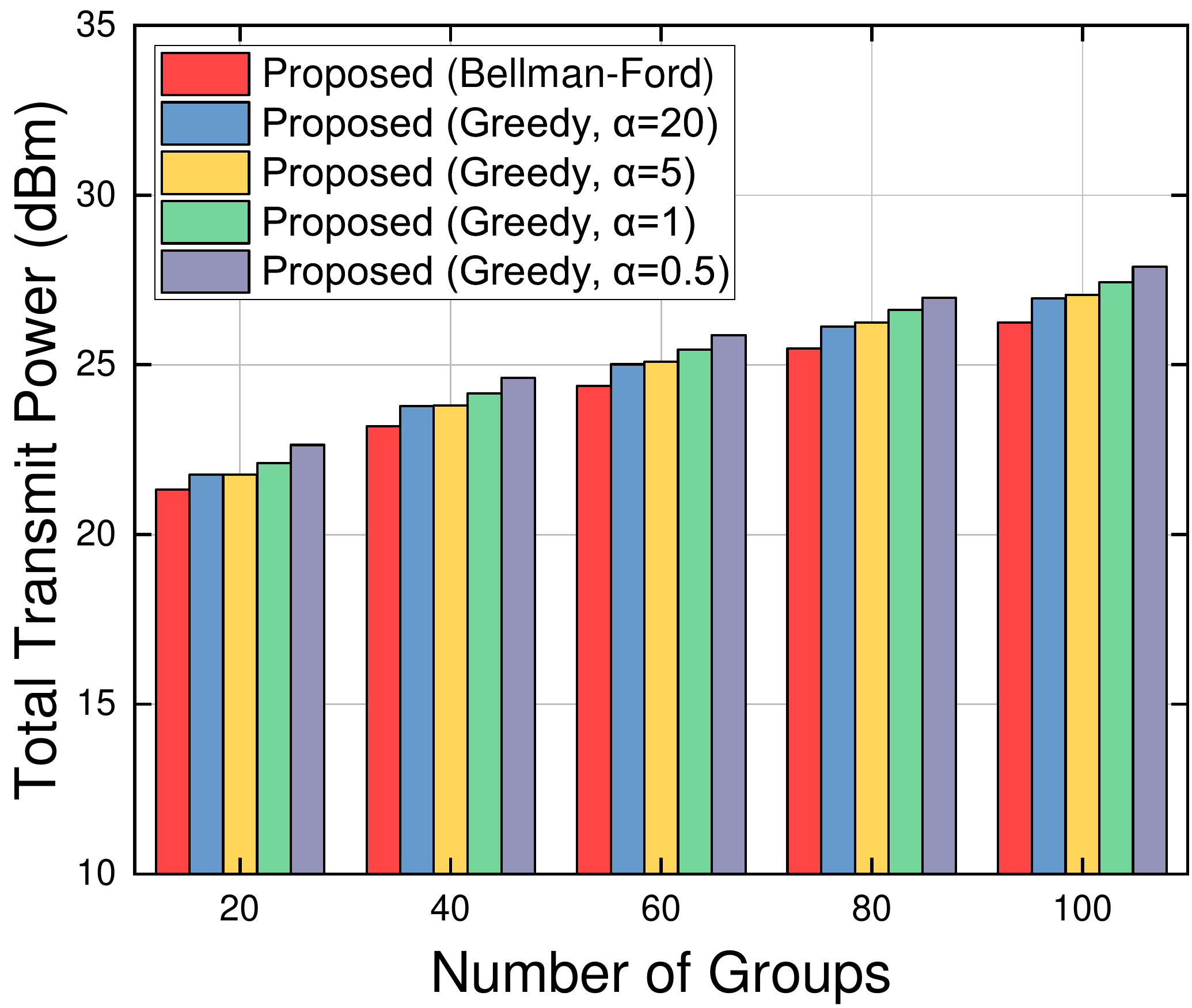}
	}
	\vspace{.1in}
	\subfigure[]{
		\label{fig:3}
		\includegraphics[width=1.634in]{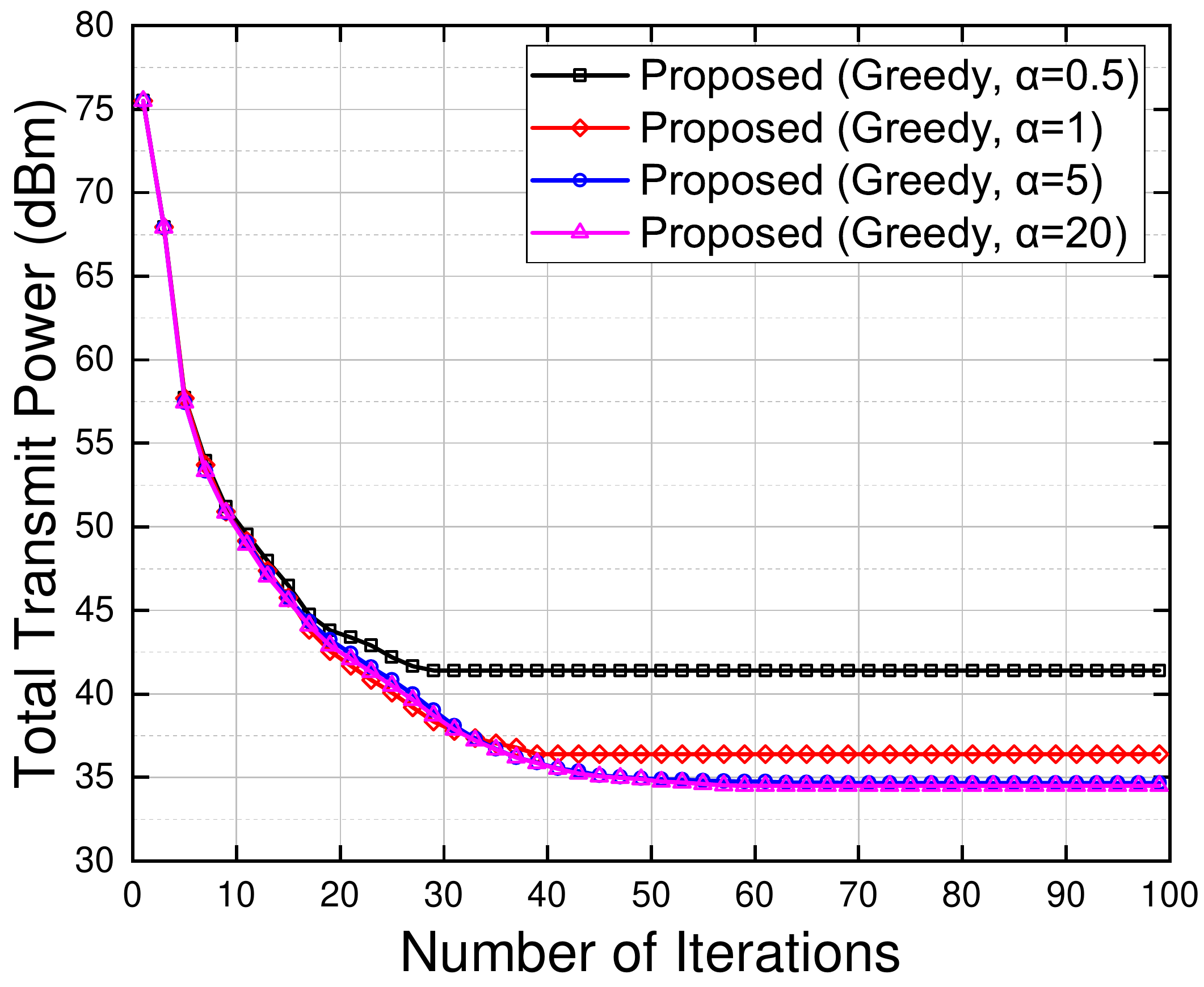}
	}
	\caption{Impacts of $ \alpha $. (a)Total Transmit Power vs. Number of groups;(b)Total Transmit Power vs. Number of Iterations}
	\label{figa}
\end{figure}

Fig. \ref{fig:6} shows the total transmit power allocated to all users with $G=[20, 40,\cdots,100]$ and the number of users is set to be two times as the number of groups. As expected, the power consumption increases with the number of groups for all the two algorithms. In addition, the power consumption decreases slowly with the increase of $\alpha$ and almost changes no more when $\alpha>=5$. Fig. \ref{fig:3} shows the total transmit power for all users with  iterations, where $ G=100 $, $ N=300 $. We can see that the convergence rate increases with decreasing $ \alpha $ and almost remains constant when $\alpha>=5$. In addition, the total transmit power declines rapidly at the beginning and tends to a stable value after finite iterations.

%
%


\subsection{Performance Comparison}

The proposed algorithms are compared with existing two reference user grouping strategies. One is assigning users one by one according to their preference lists (``User Preference'') \cite{Di1}. The other is grouping users by Gale-Shapley algorithm(``Gale-Shapley'') \cite{Xu2018}. In the simulations, the number of users is set to an integer multiple of the number of groups. This is because in the reference strategies, the number of users in each group should be equal.

\begin{figure}
	\centering
	\subfigure[]{
		\label{fig:1}
		\includegraphics[width=1.616in]{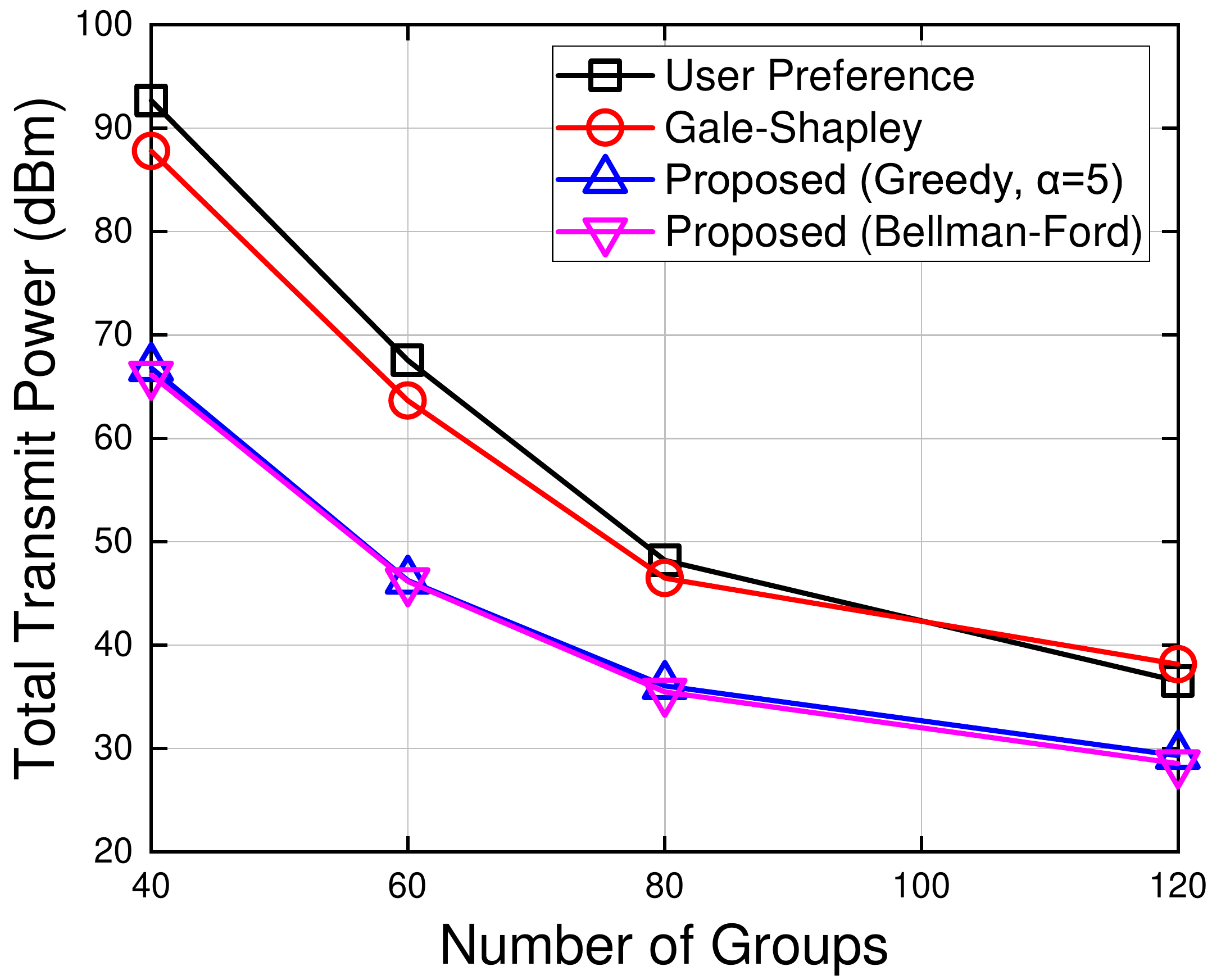}
	}
	\subfigure[]{
		\label{fig:2}
		\includegraphics[width=1.616in]{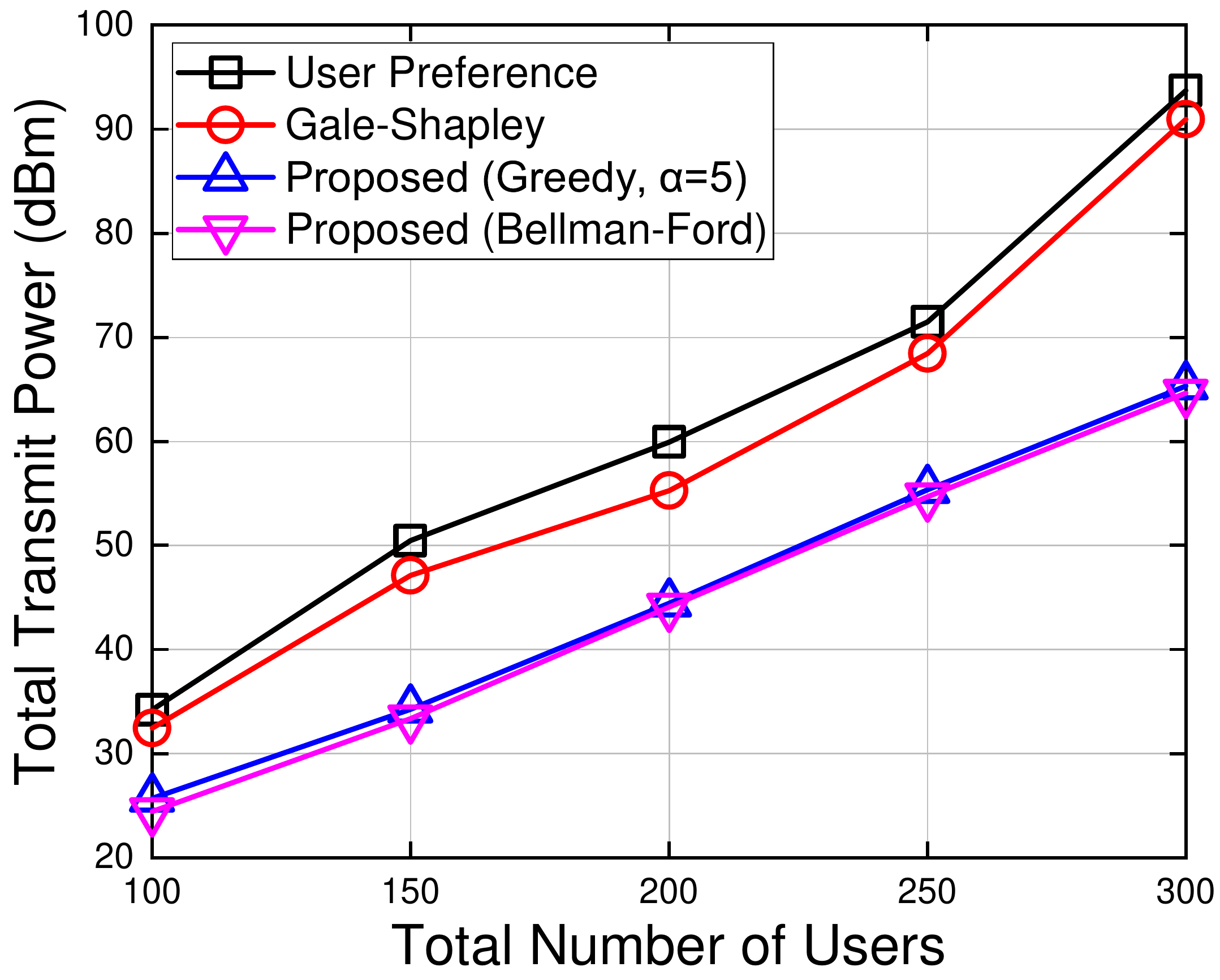}
	}
	\caption{Average total transmit power comparison among different strategies. (a)Total Transmit Power vs. Number of groups (240 users in total);(b)Total Transmit Power vs. Number of users (50 groups in total)}
	\label{figb}
\end{figure}

Fig. \ref{fig:1} shows the total transmit power allocated to all users with $G=[40, 60,80,120]$ and 240 users in total. As expected, the power consumption decreases with the number of groups for all the four strategies. In addition, the proposed algorithms outperform the reference strategies. The reason is that the reference strategies ignore the impact of interference among users on user grouping. Note that performance of Proposed-Greedy($\alpha=5$) approaches that of Proposed-BellmanFord.

Fig. \ref{fig:2} shows the total transmit power with varied number of users and 50 groups in total. As expected, the power consumption increases rapidly with the number of users for these four strategies. This is because interference among users in the same group will rapidly increases as the number of users in each group increases. This result confirms that it's unrealistic to implement NOMA among all users simultaneously. Furthermore, the proposed strategies outperform the reference strategies. In addition, as the number of users increases, the total transmit power difference between the proposed algorithms and the reference strategies increases. This is because there will be much room for power reduction by user grouping as the number of users in each group increases, and the proposed algorithms can reduce the power consumption more effectively with consideration of interference and QoS constraints. 



\begin{figure}[ht]
	\centering
	\includegraphics[width=3.4in]{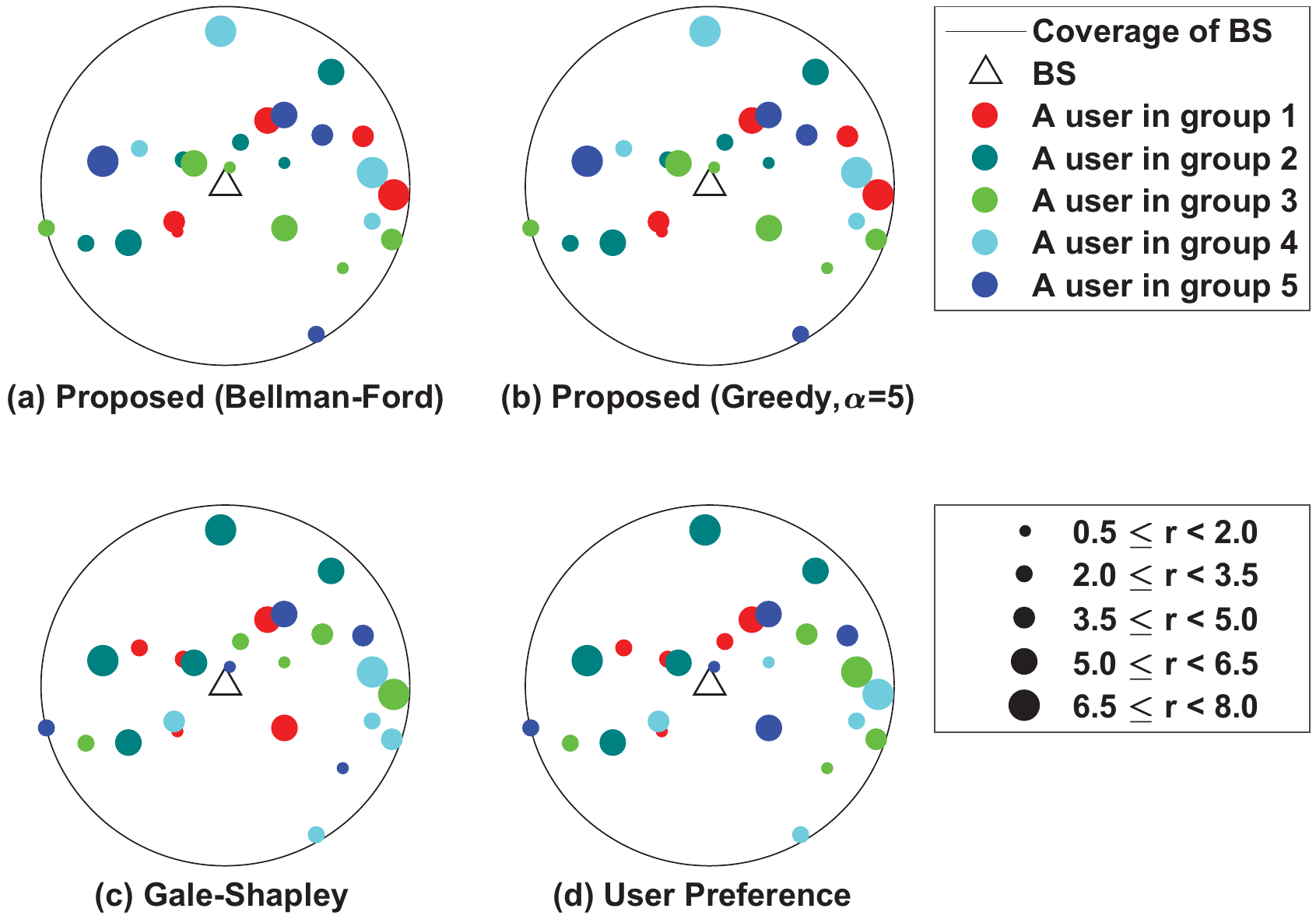}
	\caption{The distribution of the users for different strategies}
	\label{fig:4}
\end{figure}

As mentioned in Section II, there are two important principles in user grouping to reduce interference among the users in the same group. One is the users with high target data rate should be avoided to be assigned into the same group. The other one is the users with poor channel condition should be avoided to be assigned into the same group. Much more power will be consumed if one of these principles is unsatisfied. To show the superiority of the proposed algorithms, Fig. \ref{fig:4} shows the distribution of users ,where the number of users and the number of groups are set to 25 and 5, respectively. Each dot represents a user. The dots in the same color denote the users in the same group, the distance between user and the BS represents the channel condition, and the size of dot represent the size of this user's target data rate. We can see from Fig. \ref{fig:4}(c) that there are four users close to the edge of the BS range in group $4$, which leads to severe interference among users in group $4$. In addition, both in Fig. \ref{fig:4}(c) and Fig. \ref{fig:4}(d), there are five users whose target data rates are more than 5 bps/Hz in group $5$, which also make interference among users in group $4$ severe. However, such interference is avoided in the proposed algorithms as \ref{fig:4}(a) and \ref{fig:4}(b) show. In addition, the performance of Proposed-Greedy($\alpha=5$) is similar to that of Proposed-BellmanFord while the computational complexity is significantly reduced.





\section{Conclusion}
In this paper, we solved the user grouping problem with QoS constraints by graph theory. First, we defined the PCE function for each user to represent the power consumption and interference introduced by this user when it joints a group. It has been proved that the difference of the PCE function of a user could be used to indicate the difference of total power consumption when it changes its grouping strategy. Based on the definition of PCE, we constructed a directed graph. Then, in multi-user scenarios, we introduced the conception of \emph{exchange league} to judge whether user grouping can be adjusted to reduce the power consumption. As the \emph{exchange league} has been proved to be equivalent to the negative loop in the graph, we converted the user grouping problem into the problem of searching specific negative loops with all users in different groups. Bellman-Ford algorithm has been extended to find these negative loops. Furthermore, a greedy suboptimal algorithm has been proposed to approach the solution with polynomial time. Finally, simulation results have demonstrated the advantages of the proposed algorithms compared with existing grouping strategies under different scenarios.




\bibliographystyle{IEEEtran}
\bibliography{bibtex}
\end{document}